\documentclass[twocolumn]{aastex62}
\pdfoutput=1 
\usepackage{amsmath,amstext}
\usepackage{nicefrac}
\usepackage[T1]{fontenc}
\usepackage{apjfonts} 
\usepackage[figure,figure*]{hypcap}
\usepackage{float}

\renewcommand{\thefootnote}{\textit{\alph{footnote}}} 

\shorttitle{Discovery of a Bright QSO}
\shortauthors{Jeram et al.}

\begin{document}

\def\qsoname{2MASS J13260399+7023462}
\def\solmass{M$_{\odot}$}
\def\sollum{L$_{\odot}$}
\def\redshift{2.889}
\def\redshifterr{0.002}
\def\redshiftAnderr{$2.889 \pm 0.002$}
\def\fnuFourteenFifty{$0.89 \pm 0.15$ mJy}
\def\fnu2500{$1.31 \pm 0.15$ mJy}
\def\mAB1450{$16.6 \pm 0.2$}
\def\MAB1450{$-28.9 \pm 0.2$}
\def\monolum{($1.2 \pm 0.2$) $\times$ 10$^{48}$ ergs s$^{-1}$}
\def\Lbol{($4.7 \pm 0.9$) $\times$ 10$^{41}$ Watts}
\def\Lbolergs{($4.7 \pm 0.9$) $\times$ 10$^{48}$ erg s$^{-1}$}
\def\LbolLsol{($1.2 \pm 0.2$) $\times$ 10$^{15}$ \sollum}
\def\Mbol{$-33.0 \pm 0.2$}
\def\accrate{$557 \pm 100$ \solmass yr$^{-1}$}
\def\eddrate{$1.3 \pm 0.3$}
\def\CIVblueshift{$1740 \pm 50 \text{ km s}^{-1}$}
\def\CIVFWHM{$8000 \pm 150 \text{ km s}^{-1}$}
\def\CIVFWHMcorr{$6000 \pm 360 \text{ km s}^{-1}$}
\def\bhmass{($2.7 \pm 0.4$) $\times$ 10$^{10}$ \solmass}
\def\radioopticalratio{$1.9 \pm 0.2$}
\def\QLFestimate{37}

\title{An Extremely Bright QSO at $z=2.89$}
\author[0000-0003-1071-144X]{Sarik Jeram}
\affiliation{Department of Astronomy, University of Florida, Bryant Space Science Center, Gainesville, FL 32611, USA}
\author{Anthony Gonzalez}
\affiliation{Department of Astronomy, University of Florida, Bryant Space Science Center, Gainesville, FL 32611, USA}
\author{Stephen Eikenberry}

\affiliation{Department of Astronomy, University of Florida, Bryant Space Science Center, Gainesville, FL 32611, USA}
\affiliation{Department of Physics, University of Florida, New Physics Building, Gainesville, FL 32611, USA}
\author{Daniel Stern}
\affiliation{Jet Propulsion Laboratory, California Institute of Technology, 4800 Oak Grove Drive, Pasadena, CA 91109, USA}
\author{Claudia Lucia Mendes de Oliveira}
\author{Lilianne Mariko Izuti Nakazono}
\affiliation{Departamento de Astronomia, Instituto de Astronomia, Geof\'isica e Ci\^encias Atmosf\'ericas da USP, Rua do Mat\~ao 1226, Cidade Universit\'aria, 05508-090, S\~ao Paulo, Brazil}
\author{Kendall Ackley}
\affiliation{OzGrav, School of Physics \& Astronomy, Monash University, Clayton 3800, Victoria, Australia}


\begin{abstract}
We report the discovery and confirmation of a bright quasi-stellar object (QSO), \qsoname{},  at $z=\redshift{}$. This QSO is the first spectroscopically confirmed candidate from an ongoing search using the combination of Gaia and WISE photometry to identify bright QSOs at $z>2$, the redshift regime for which the Lyman-$\alpha$ forest is accessible with ground-based facilities. With a Gaia apparent magnitude $G=16.07$, \qsoname{} is one of the brightest QSOs known at $z>2$, with only 15 currently known brighter QSOs. Given its inferred $M_{1450,AB}$ magnitude and redshift, it is among the most luminous objects in the Universe; the inferred black hole mass and corresponding Eddington ratio are \bhmass{} and \eddrate{}, respectively. Follow-up {\it Hubble} observations confirm it is not gravitationally lensed.
\end{abstract}


\section{Introduction}
The epoch from reionization to the era of peak star formation ($2 \lesssim z \lesssim 7$, \citealt{sfh}) is a time of rapid galaxy growth and assembly \citep{galaxygrowth}, as well as commensurate growth of supermassive black holes (SMBHs) in galaxies. Indeed, one of the outstanding challenges in galaxy formation is explaining the existence of the most massive SMBHs at $z \gtrsim 7$ \citep{bhev, banados2018}, which are observed as quasi-stellar objects (QSOs).  More generally, evolution of the bright end of the quasar luminosity function (QLF) provides an observational benchmark which must be reproduced by theoretical models \citep{obsQLF}.


Beyond evolution of the QLF, ultraluminous QSOs are also of interest as physical probes of the Universe. Bright, lensed QSOs can be used for time delay measurements to constrain the Hubble constant $H_0$, as first described by \cite{refsdaltimedelay}. Bright, unlensed QSOs can serve as backlights for studying the Lyman-$\alpha$ forest \citep{BOSSlaf} and as tools for detecting cosmological redshift drift, as described in \cite{loeb98}. The latter of these is an extension of a test first proposed by \citet{sandage}, in which one measures the change in the expansion velocity of the Universe via the Lyman-$\alpha$ forest absorption lines superposed on the QSO spectrum. \citet{liske} demonstrated that with a small sample of bright QSOs and a 30-meter class telescope, direct measurements of acceleration in the expansion are possible. Ground-based Lyman-$\alpha$ forest studies and cosmological redshift drift require QSOs at $z>2$ so that Lyman-$\alpha$ lies in the optical window.

Significant progress has been made in recent years in quantifying the bright end of the QLF \citep{lutzQLF, richardsQLF, rossQLF, elqsS}, including development of machine learning techniques and utilization of wide-area optical surveys like Pan-STARRS \citep{elqsN, elqsPS, elqsS}. This census remains incomplete, as until recently all-sky optical data was unavailable and discrimination of QSOs from faint stars can be challenging. The combination of all-sky data from the Gaia and WISE missions offers a uniquely powerful means of completing this census. Our team has initiated a search to identify the brightest $z>2$ quasars using these data sets. In this work, we report on the discovery of the first optically bright, ultraluminous QSO identified in this search. For pure luminosity evolution, the QLF is best described by a broken double power law consisting of a faint-end slope, a bright-end slope, the break magnitude between these slopes, and the overall density normalization \citep{boyle88, pei95, boyle00}. As shown in \citet{mcgreer13}, the break magnitude evolves strongly in the redshift range $2.8 < z < 4.5$, affecting both the bright-end slope and density normalization \citep{elqsS}. Discoveries from high-z QSO searches such as the one presented here will provide additional constraints on the bright-end slope and break magnitude.

\section{Search Overview}
\subsection{Data}
We use data from the AllWISE and Gaia DR2 source catalogs to identify QSO candidates \citep{wise, Gaia}. From the Gaia DR2 source catalog, accessed via the Gaia Archive (\url{http://gea.esac.esa.int/archive}), we use Gaia $G$, $G_{\text{BP}}$, and $G_{\text{RP}}$ magnitudes, parallaxes, and proper motions and use the pre-cross-matched ``gaiadr2.allwise\_best\_neighbor" table list to obtain profile-fit magnitudes (w?mpro) for bands 1 through 4 from the AllWISE source catalog.

\subsection{Method}


The method we use to discover this object employs a standard optical/infrared color selection to isolate QSOs from stars in color-color space. A detailed description of the detection method used to discover this quasar will appear in an upcoming paper. Briefly, we start by using the composite QSO spectrum from \cite{hcqso}, which extends to the mid-infrared, to compute the expected colors of a QSO as a function of redshift. We also compute colors for synthetic stellar spectra from \cite{starspectra} for comparison.  We then construct a set of selection criteria based upon the $G_{\text{RP}}-$W1 and W1$-$W2 colors designed to optimize selection of $z\ga2$ QSOs and minimize stellar contamination. Because the focus of our work is the identification of the brightest QSOs at this epoch, we initially restrict our attention to objects with Gaia $G < 16.5$. Historically, the limitation with this approach has been the similarity of QSO colors to those of main-sequence stars. Coupled with the color selection, we use Gaia DR2 to reject any stars with 3-$\sigma$ detections of proper motions and parallaxes.

As a test of the completeness resulting from this approach, we checked the efficiency with which we recover known high-redshift QSOs. Of the 282 known $z>2$ QSOs with $G_{BP}<17.5$, our proper motion, parallax, and color selection criteria recover 90\%. We prioritized initial follow-up based upon proximity of a candidate to the the track in color space of the QSO template. We selected \qsoname{}, which has a cross-identification in Gaia DR2 of 1685441172355283328, as the first target for spectroscopic follow-up as it lies nearly atop the \citet{hcqso} track and is proximate to a known QSO with $z = 2.55$.

\section{Spectroscopic Confirmation and SED Analysis}
\begin{figure}[b]
    \centering
    \includegraphics[height=8.5cm,width=8.5cm]{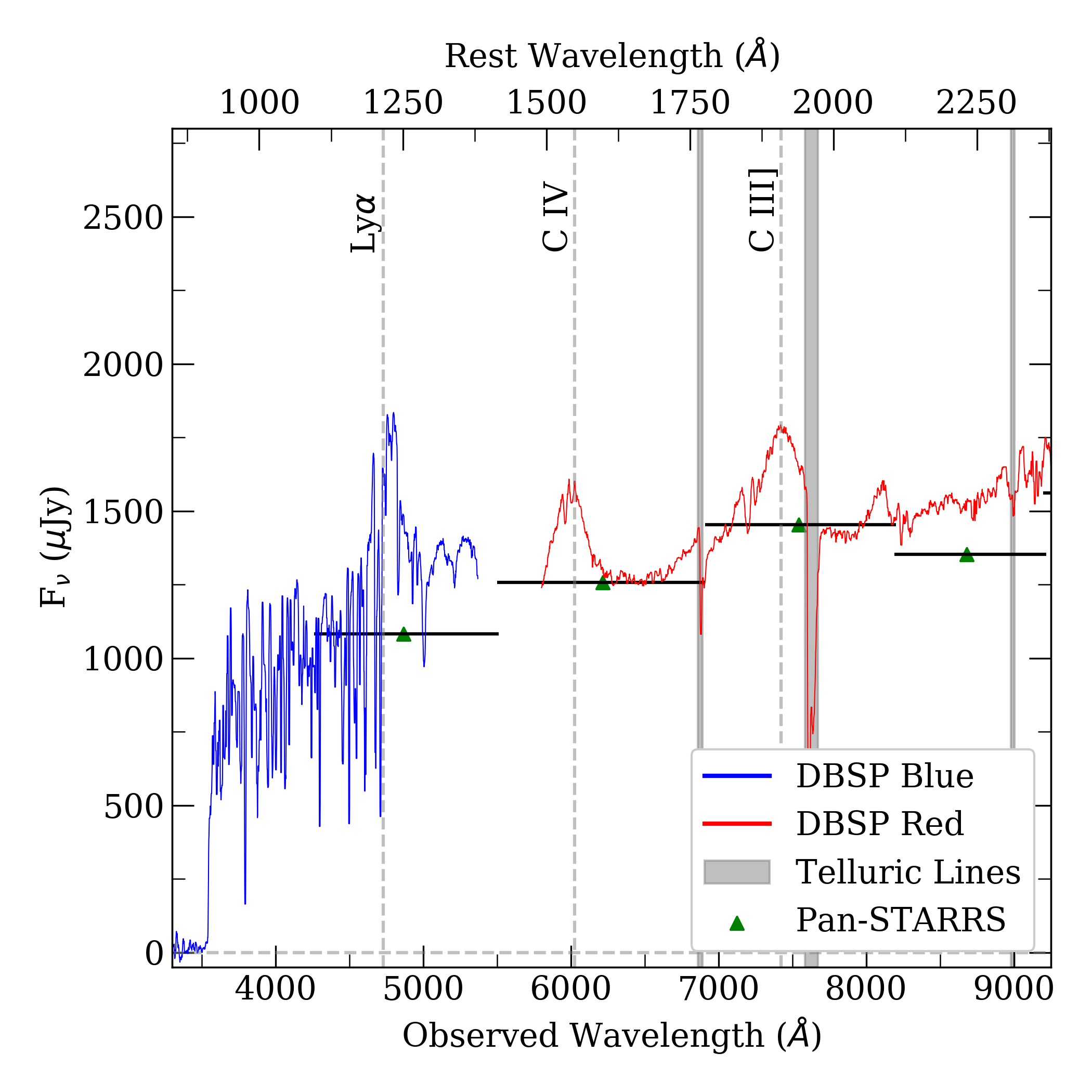}
    \caption{The optical spectrum of \qsoname{} taken with DBSP at Palomar Observatory shows several expected emission lines at $z=\redshift{}$ along with Lyman-$\alpha$ forest absorption lines. Telluric absorption bands are highlighted in gray.}
    \label{spectra}
\end{figure}

We used the Double Spectrograph (DBSP; \citealt{dbsp}) at the 200-inch Hale telescope at Palomar Observatory to observe \qsoname{} on the night of UT 2018 June 6. We obtained a single 300 second exposure using DBSP on a night with good seeing conditions but with thin cloud observed at sunrise. A 1.5'' wide slit was used with a 5500\AA{} dichroic beam splitter, a 600 $\ell$ mm$^{-1}$ grating in the blue channel ($\lambda_{\text{blaze}}=4000$\AA{}; spectral resolving power $R \equiv \nicefrac{\lambda}{\Delta \lambda} \sim 1200$), and a 316 $\ell$ mm$^{-1}$ grating in the red channel ($\lambda_{\text{blaze}}=7500$\AA{}; spectral resolving power $R \sim 1800$). The data were processed using standard routines found in the IRAF software package and observations were calibrated with Feige 34 (white dwarf) and BD+28 4211 (subdwarf O-star). The observed spectra are shown in Figure \ref{spectra}. The presence of the Lyman-$\alpha$ peak and forest along with CIV and CIII] broad emission lines confirms that this object is a QSO. 

\begin{figure}
    \centering
    \includegraphics[height=8.5cm,width=8.5cm]{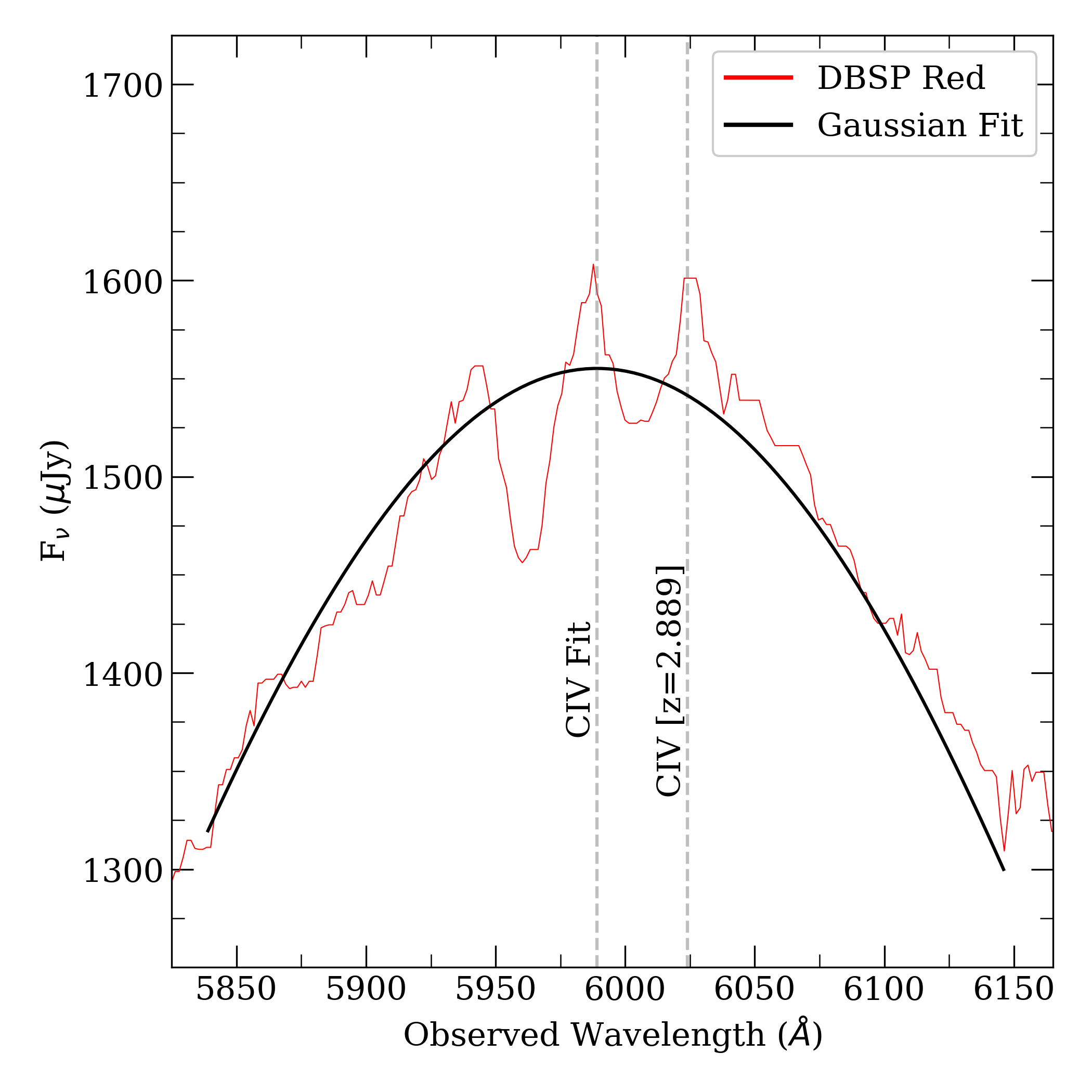}
    \caption{CIV line fit with a Gaussian profile to determine the velocity offset relative to the QSO rest frame. The CIV profile is blueshifted relative to the CIII]-defined QSO rest frame by \CIVblueshift{}.}
    \label{civfit}
\end{figure}

\begin{figure}
    \centering
    \includegraphics[height=8.5cm,width=8.5cm]{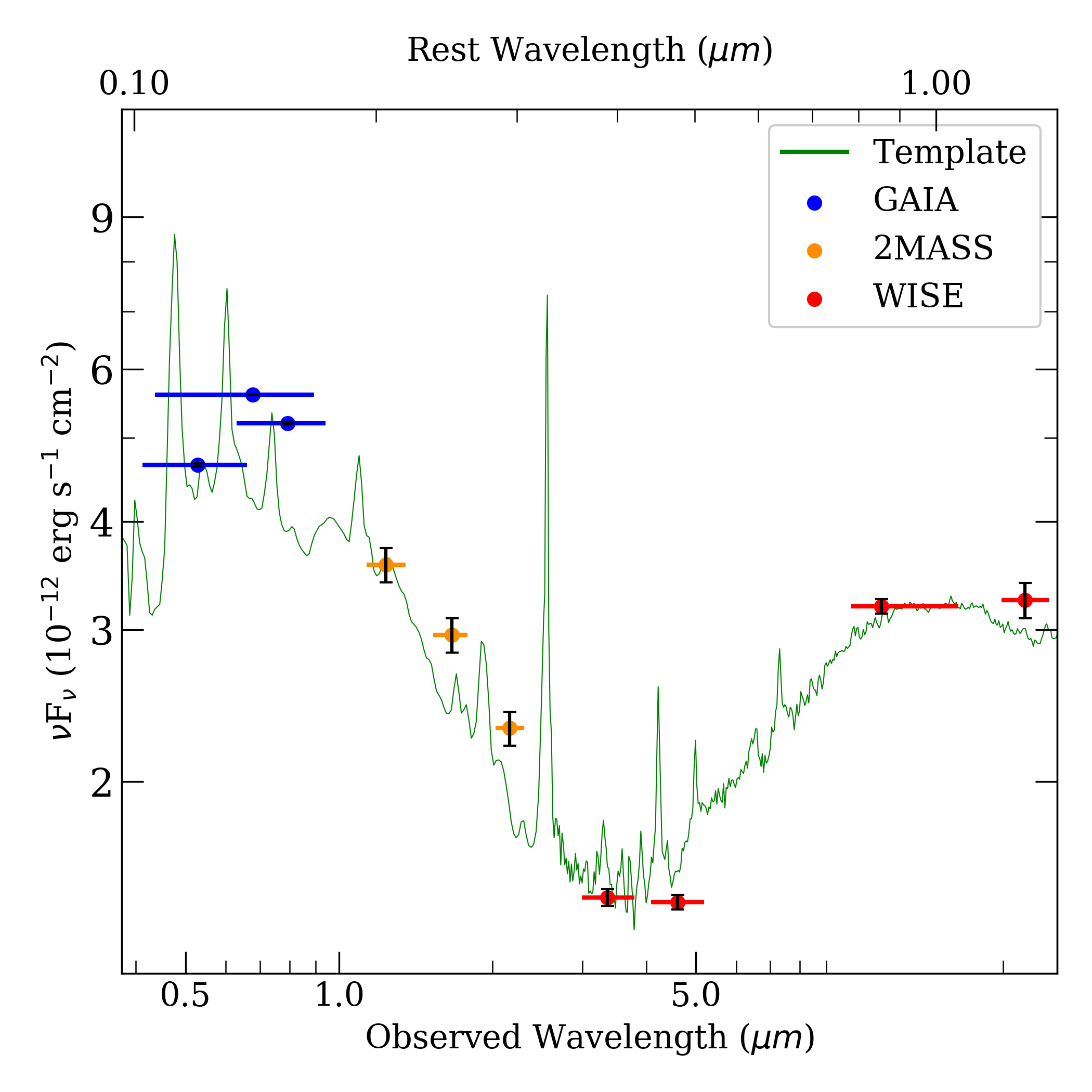}
    \caption{The SED of \qsoname{} with the composite spectrum of \citet{hcqso} overplotted for comparison. The passbands are the same as those listed in Table \ref{qsotable} and are represented by the horizontal line for each point. The flux values are calculated using the zeropoints from the SVO Filter Profile Service.}
    \label{sed}
\end{figure}

To determine the redshift, we fit a Gaussian profile to the CIII] feature in the spectrum and compare it to its known rest-frame wavelength (1909\AA{}). We avoid using the Lyman-$\alpha$ line for this purpose as the multitude of absorption features bluewards of it prevent us from obtaining a reliable fit. We also avoid using the CIV feature as it is often blueshifted due to outflows \citep{richardsBlueshift, netzerOutflow}. We determine the redshift solely from the observed CIII] broad line, taking care to avoid the telluric absorption feature on the red side of the line, and obtain $z = $ \redshiftAnderr{}. Using this redshift, we calculate the blueshift of the CIV line to be \CIVblueshift{} in the rest frame of the system after fitting a Gaussian profile, the FWHM of which is \CIVFWHM{}; the uncertainty of these measurements is determined by varying the wavelength range over which the line profile is fit. The line fit and blueshifting are shown in Figure \ref{civfit}.

Using the zeropoints specified in the Spanish Virtual Observatory (SVO) Filter Profile Service  \citep{svo}, we calculate fluxes for each of the passbands listed in Table \ref{qsotable}. We then redshift the composite template and scale it to match 2MASS {\it J}-band photometry, the bluest passband at which the spectrum is devoid of strong emission lines, and overplot the fluxes to create the SED shown in Figure \ref{sed}.


\cite{vandenberk} show that for a QSO observed in the SDSS $r$-band (the filter corresponding to rest-frame 1450\AA{} at $z$ = \redshift), brightness fluctuations on the order of $\sim$0.2 magnitudes can be expected. In addition to this, \cite{variability} show that the average characteristic timescale for QSO variability in the rest frame is approximately 2 years -- this is of particular relevance given that the most recent 2MASS, AllWISE, and Gaia datasets were released in the years 2003, 2013, and 2018, respectively.

\begin{figure}
    \centering
    \includegraphics[height=12cm,width=8.5cm]{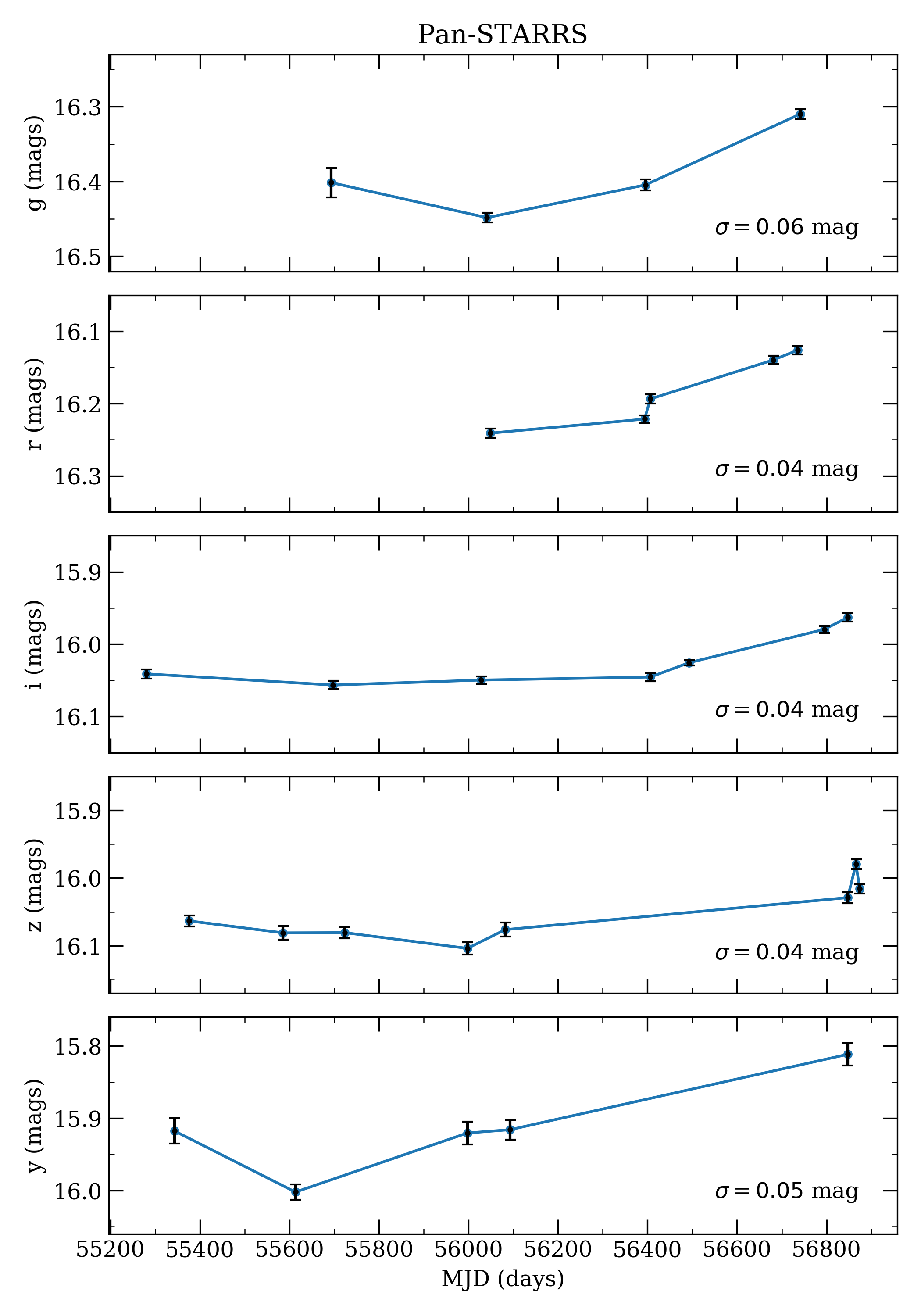}
    \caption{The Pan-STARRS DR1 light curve of \qsoname{}, taken over a $\sim$4 year span from 2010$-$2014. These data show a net brightening, as is readily observed in the $y$-band light curve.}
     \label{ps_var}
\end{figure}

To investigate the variability of 2MASS J13260399+ 7023462, we compile and plot single epoch Pan-STARRS data, the results of which we show in Figure \ref{ps_var}. The light curve exhibits variation on a scale consistent with what is suggested by \cite{vandenberk}, lending credibility to the hypothesis that some portion of the deviations from the composite are a direct consequence of the variability itself. The multi-year Pan-STARRS light curve shows a net increase in brightness from the start of observations, with the $g$ and $y$-band light curves in particular showing a prominent brightening of $\sim$0.2 magnitudes. The light curve shows variations that far exceed the photometric uncertainties provided in Table \ref{qsotable}. We note that a brightness increase in one set of passbands does not necessitate a similar brightness increase in other passbands. Furthermore, the dust content of any one QSO is not guaranteed to be similar to the dust content the composite template assumes.

Although we do not utilize variability information in the current implementation of our search method, it can be used to distinguish QSOs from contaminants. \cite{qsoVariabilitySelection} used intrinsic variability to select quasars with high completeness and purity comparable to existing color selection methods for the redshift range $2.5 < z < 3$.


\begin{table}
\centering
\caption{Summary information about \qsoname{}. All magnitudes reported in the photometric information section are in the Vega system. The luminosity distance and absolute magnitude at 1450\AA{} were calculated using the parameters discussed in Section \ref{results}.}
\label{qsotable}
\begin{tabular}{l r}
\hline
\multicolumn{2}{c}{Positional Info}\\ \hline

\multicolumn{1}{l}{DR2 Source ID} & \multicolumn{1}{r}{1685441172355283328}\\
\multicolumn{1}{l}{RA (deg)} & \multicolumn{1}{r}{201.5166465}\\
\multicolumn{1}{l}{DEC (deg)} & \multicolumn{1}{r}{70.3961983}\\
\multicolumn{1}{l}{b$_{gal}$ (deg)} & \multicolumn{1}{r}{118.7285419}\\
\multicolumn{1}{l}{l$_{gal}$ (deg)} & \multicolumn{1}{r}{46.4484052}\\

\hline
\multicolumn{2}{c}{Photometric Info}\\ \hline

\multicolumn{1}{l}{GALEX FUV} & \multicolumn{1}{r}{undetected}\\
\multicolumn{1}{l}{GALEX NUV} & \multicolumn{1}{r}{undetected}\\
\multicolumn{1}{l}{Gaia $G$} & \multicolumn{1}{r}{16.070 $\pm$ 0.001}\\
\multicolumn{1}{l}{Gaia $G_\text{BP}$} & \multicolumn{1}{r}{16.348 $\pm$ 0.005}\\
\multicolumn{1}{l}{Gaia $G_\text{RP}$} & \multicolumn{1}{r}{15.643 $\pm$ 0.004}\\
\multicolumn{1}{l}{Pan-STARRS $g$} & \multicolumn{1}{r}{16.393 $\pm$ 0.013}\\
\multicolumn{1}{l}{Pan-STARRS $r$} & \multicolumn{1}{r}{15.997 $\pm$ 0.008}\\
\multicolumn{1}{l}{Pan-STARRS $i$} & \multicolumn{1}{r}{15.625 $\pm$ 0.009}\\
\multicolumn{1}{l}{Pan-STARRS $z$} & \multicolumn{1}{r}{15.563 $\pm$ 0.006}\\
\multicolumn{1}{l}{Pan-STARRS $y$} & \multicolumn{1}{r}{15.375 $\pm$ 0.016}\\
\multicolumn{1}{l}{2MASS $J$} & \multicolumn{1}{r}{15.089 $\pm$ 0.050}\\
\multicolumn{1}{l}{2MASS $H$} & \multicolumn{1}{r}{14.489 $\pm$ 0.050}\\
\multicolumn{1}{l}{2MASS $K_{s}$} & \multicolumn{1}{r}{14.009 $\pm$ 0.049}\\
\multicolumn{1}{l}{WISE W1} & \multicolumn{1}{r}{13.187 $\pm$ 0.024}\\
\multicolumn{1}{l}{WISE W2} & \multicolumn{1}{r}{12.217 $\pm$ 0.021}\\
\multicolumn{1}{l}{WISE W3} & \multicolumn{1}{r}{8.527 $\pm$ 0.021}\\
\multicolumn{1}{l}{WISE W4} & \multicolumn{1}{r}{6.364 $\pm$ 0.051}\\

\hline
\multicolumn{2}{c}{Miscellanous Info}\\ \hline

\multicolumn{1}{l}{$z$} & \multicolumn{1}{r}{\redshift{} $\pm$ \redshifterr{}}\\
\multicolumn{1}{l}{$D_L$} & \multicolumn{1}{r}{24.76 Gpc}\\
\multicolumn{1}{l}{$m_{1450, AB}$} & \multicolumn{1}{r}{\mAB1450{}}\\
\multicolumn{1}{l}{$M_{1450, AB}$} & \multicolumn{1}{r}{\MAB1450{}}\\
\multicolumn{1}{l}{$M_\text{bol}$} & \multicolumn{1}{r}{\Mbol{}}\\
\multicolumn{1}{l}{$L_{\text{bol}}$} & \multicolumn{1}{r}{\LbolLsol}\\
\multicolumn{1}{l}{$M_{BH}$} & \multicolumn{1}{r}{\bhmass}\\
\multicolumn{1}{l}{$\lambda_{\text{Edd.}}$} & \multicolumn{1}{r}{1.3 $\pm$ 0.3}\\
\hline
\end{tabular}
\end{table}

\section{Results} \label{results}
A standard measure of QSO luminosity is the absolute magnitude at 1450\AA{} (M$_{1450}$) due to the lack of emission features that overlap this wavelength. We estimate this quantity using the \citet{hcqso} composite after scaling it to match the Gaia photometry for this QSO. We find F$_\nu$ at 5623\AA{}, the observed wavelength of rest-frame 1450\AA{} at $z$ = \redshift{}, to be F$_{\nu,1450} =$ \fnuFourteenFifty{} from which we calculate m$_{1450} =$ \mAB1450{}. Using the cosmology specified in  \citet{planck2018} for a flat Universe, the implied absolute magnitude is M$_{1450} =$ \MAB1450. The most luminous, unlensed QSO at this epoch, HS 1946+7658, is of a comparable absolute magnitude with M$_{1450} = -29.2$, indicating that \qsoname{} is one of the most luminous QSOs known. Using the aforementioned F$_{\nu, 1450}$ and luminosity distance, we infer a monochromatic luminosity of $\nu L_{\nu,1450} =$ \monolum{}. 

We next compute the bolometric luminosity for this system. We find $L_{\text{bol}} =$ \Lbolergs{}, equivalent to \LbolLsol, assuming a bolometric correction of 3.8 at 1450\AA{} as in \citet{wolf}. This luminosity corresponds to a bolometric magnitude for this QSO of \Mbol{}. We measure the FWHM of the CIV line in the spectrum to be \CIVFWHM{}, from which we obtain a corrected FWHM of \CIVFWHMcorr{} using Equation 4 from \cite{coatmanbhmass} to correct for non-virial CIV emission. We estimate the monochromatic luminosity at 1350\AA{} from our scaled composite to be 1.5 $\times$ 10$^{48}$ erg s$^{-1}$ at 1350\AA{}. Using these quantities with Equation 6 from \cite{coatmanbhmass}, we obtain a value of \bhmass{} for the black hole mass. For comparison, the most massive black hole currently known is TON 618 with a mass of 6.6 $\times$ 10$^{10}$ \solmass, indicating that the black hole powering \qsoname{} is among the most massive currently known \citep{ton618mass}. Using $L_{\text{bol}}$ and $L_{\text{Edd.}} = 1.3 \times 10^{38} \big(\frac{M_{\text{BH}}}{M_{\odot}}\big)$ erg s$^{-1}$ from \citet{rybickilightman}, we determine the Eddington ratio, $\lambda_{\text{Edd.}} = \nicefrac{L_{\text{bol}}}{L_{\text{Edd.}}}$ to be \eddrate{}, which indicates \qsoname{} is currently undergoing accretion near its Eddington limit.
\clearpage



\begin{figure}[]
    \centering
    \includegraphics[height=8.5cm,width=8.5cm]{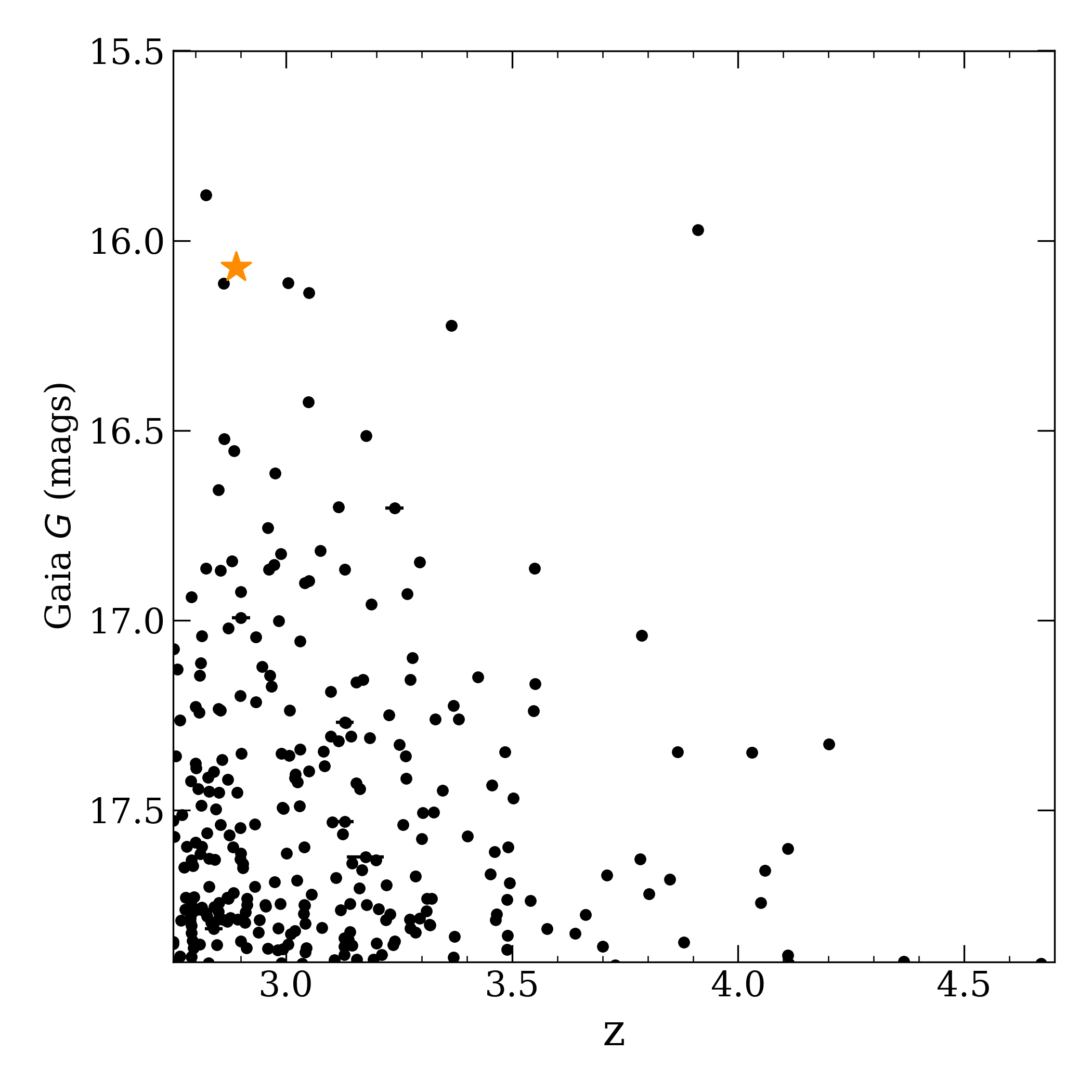}
    \caption{The brightness of \qsoname{} relative to $z>2$ QSOs from SIMBAD which have been cross-matched with Gaia DR2 shows it to be exceptionally luminous -- it is the third brightest QSO in the plot at $z>2.75$.\\}
    \label{brightness_redshift}
\end{figure}

\begin{figure}
    \centering
    \includegraphics[width=\linewidth]{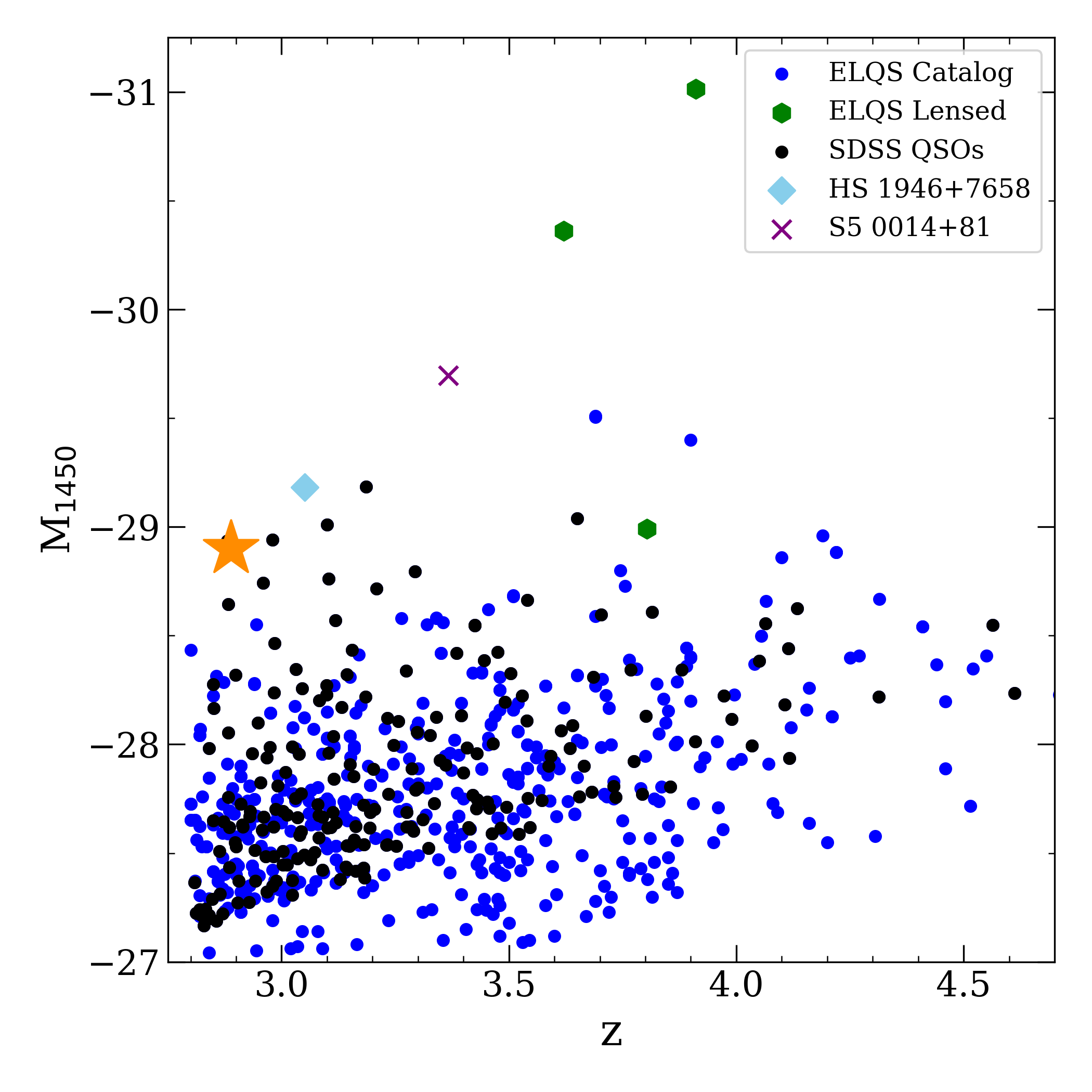}
    \caption{A comparison in the M$_{1450}$-redshift plane of \qsoname{} (gold star), the ELQS catalog (black points) \citep{elqsN, elqsPS, elqsS}, and several of the most luminous QSOs known, HS 1946+7658 (diamond) \citep{hs1946} and S5 0014+81 (cross) \citep{s5_0014}. We highlight as green diamonds the subset of the ELQS catalog that are strongly lensed.}
    \label{elqs}
\end{figure}

\subsection{\qsoname{} in Context}
To further investigate the nature of this QSO, we compare its properties with other known luminous QSOs. With $G_{\text{BP}}=16.348$ and $G=16.070$, it is the third brightest QSO at $z > 2.75$ in Figure \ref{brightness_redshift}. As such, this QSO can serve as a valuable backlight for future Lyman-alpha forest studies. To better place this QSO in context in terms of intrinsic properties, in Figure \ref{elqs} we compare it to bright QSOs at $2.8 < z < 5$ from SDSS and Pan-STARRS searches from \citet{elqsN,elqsPS,elqsS}. At this epoch, \qsoname{} currently is among the dozen most luminous known. Of those that are potentially more luminous, three are lensed QSOs (green hexagons), and hence their intrinsic luminosities are overestimated in the figure. We can further estimate the total number of brighter QSOs at this epoch (including those yet to be discovered) based on the published parameterizations of the bright end of the QLF. Using the QLF from from \citet{elqsS}, we estimate there should exist a total of \QLFestimate{} QSOs brighter than M$_{1450,AB} = -28.9$ at $2.8 < z < 4.5$. It is therefore expected there remain a number of brighter QSOs to discover at this epoch.

\subsection{Lensing}
\begin{figure}
    \centering
    \includegraphics[height=8.5cm,width=8.5cm]{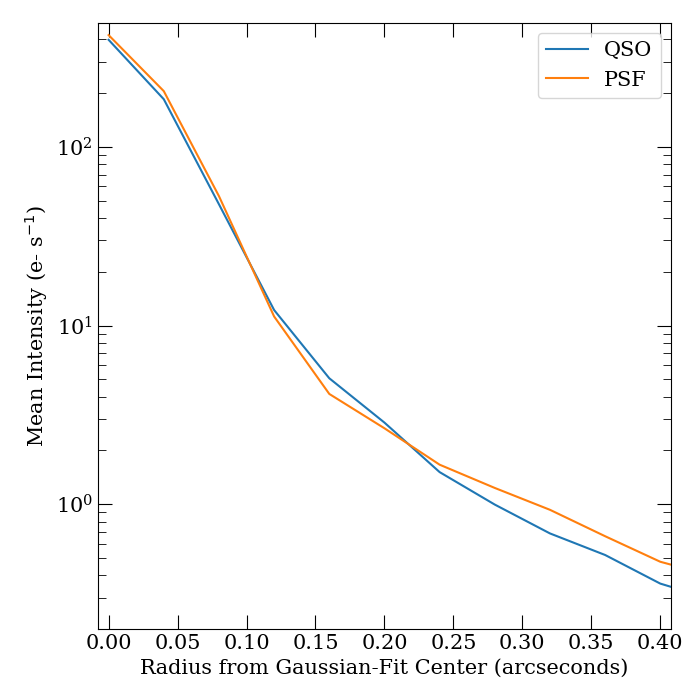}
    \caption{Radial intensity plot from the QSO and model PSF created with TinyTim for WFC3/UVIS2 F475W.}
    \label{f475w_radial}
\end{figure}

\begin{figure}[t]
    \centering
    \includegraphics[width=4.2cm]{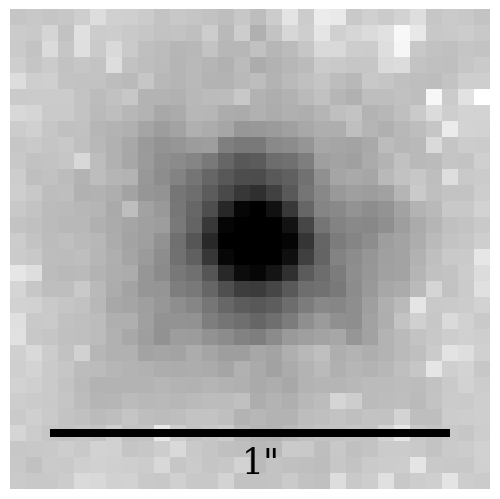}
    \includegraphics[width=4.2cm]{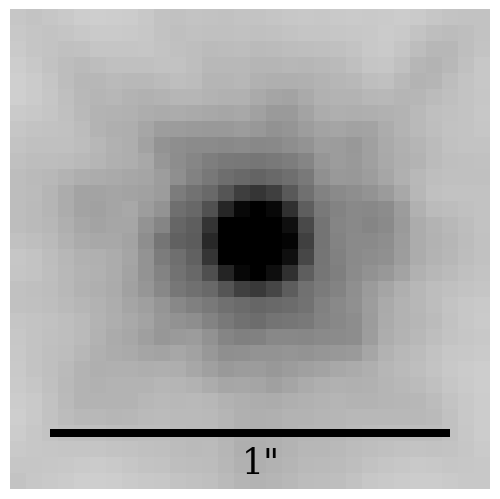}
    \label{cutouts}
    \caption{The left image shows the combined image of the {\it HST} F475W data for \qsoname{} while the right image shows the corresponding TinyTim PSF.}
\end{figure}

Given that three of the brighter QSOs at this epoch are strongly lensed, it is relevant to consider whether \qsoname{} is itself lensed, in which case its luminosity will be overestimated. To constrain any magnification due to lensing, we obtained {\it Hubble} Wide Field Camera 3 (WFC3) imaging (Program 15950) on 2020 Jan 20 in F475W within a $512\times512$ pixel subarray of UVIS2. The QSO does not resolve into multiple sources at this high resolution. To better constrain lensing, we compare the brightness profile of the QSO with the WFC3 point spread function (PSF). We use TinyTim \citep{tinytim_orig, tinytim_wfc3} to generate a model PSF for WFC3/UVIS2 F475W for each individual exposure. We then scale the model PSFs to match the amplitudes in the individual WFC3 images and combine the individual images in the same way as the actual data to derive the appropriate composite PSF. In Figure \ref{f475w_radial}, we show the comparison of the surface brightness profiles for \qsoname{} and the composite TinyTim PSF. In Figure \ref{cutouts}, we show the combined F475W image and composite TinyTim PSF.

The QSO is unresolved at HST resolution, consistent with it not being multiply imaged. We can further place an upper limit on the spatial extent by convolving the PSF with a Gaussian. A comparison of the QSO with the Gaussian-convolved PSF constrains $\sigma<0.017$ arcsec at 3-$\sigma$. If we take this to be an approximate limit on the Einstein radius, $\theta_E$, 
this corresponds to a 3-$\sigma$ mass limit on any lensing galaxy of $M < 4.0\times 10^{8}$ M$_\odot$. For halos with $M\lesssim10^{9}$ the cross section for galaxy lensing is so small (e.g. \citealt{hilbert08}) that the probability of lensing is negligible. We therefore can exclude the possibility that this source is significantly magnified by a foreground galaxy.


\subsection{Radio and X-ray Emission}
We next consider whether this source is radio-loud, and whether there is any evidence for X-ray emission. In the radio regime, \qsoname{} is outside of the FIRST \citep{first} footprint and is undetected in NVSS \citep{nvss}. Given the $\sim$2.5 mJy detection limit of NVSS and the flux density of \qsoname{} estimated from the \citet{hcqso} composite to be \fnu2500{} at 2500\AA{} (determined using the method detailed earlier in this section), we constrain the upper bound of the radio to optical flux ratio to be \radioopticalratio{}, indicating \qsoname{} is radio-quiet. 

Similarly, \qsoname{} was not detected in the ROSAT All-Sky Survey Faint Source Catalog \citep{rosatfaint}, placing an upper limit on the 0.1 to 2.4 keV flux of 5.6 $\times$ $10^{-12}$ erg cm$^{-2}$ s$^{-1}$ when assuming a photon spectrum modeled by a power law with index $\Gamma = 2$, a value typical of luminous QSOs \citep{photonindex}. Furthermore, we estimate the 2$-$10 keV X-ray luminosity using the relation described in \cite{sternXray}. We first use the scaled template to infer $\nu L_{\nu}$(6$\mu$m) $= 8.3 \times 10^{47}$ erg s$^{-1}$, from which we obtain $L$(2$-$10 keV) $= 6.6 \times 10^{45}$ erg s$^{-1}$ using the relation. The corresponding flux is then $8.9 \times 10^{-14}$ erg s$^{-1}$ cm$^{-2}$, hence this source is not expected to be detected by the extended Roentgen Survey with an Imaging Telescope Array (eROSITA) All-Sky Survey (eRASS; \cite{erosita}).

\section{Summary}
Objects like \qsoname{} are rare and historically have been difficult to find. The recent release of DR2 from  ESA's Gaia mission combined with data from WISE has enabled searches for ultraluminous, high-$z$ QSOs by allowing one to easily filter out stellar contaminants. Using a combination of parallax, proper motion, and color selection criteria, we are able to identify \qsoname{} as a bright QSO at $z = \redshift{}$. With $G=16.070$ at $z = \redshift{}$, it is the third brightest QSO known at $z > 2.75$, and one of the dozen most luminous QSOs known at $z > 2$.

With the inferred M$_{1450} = \text{\MAB1450{}}$ and bolometric magnitude of M$_{\text{bol}} = \text{\Mbol{}}$ in addition to its brightness relative other QSOs at this epoch, this QSO is extremely bright and follow-up observations with {\it Hubble} WFC3 demonstrate that this QSO is not lensed. The spectroscopic confirmation of the first QSO from our search suggests that the selection criteria we employ are able to identify interesting QSO candidates at $z$ > 2. With further refinement of this method and acquisition of spectroscopic data for additional candidates, we expect to find more bright, high-$z$ QSOs like \qsoname{}.

\section*{Acknowledgements}
Stephen Eikenberry was supported in part by a University of Florida Research Foundation Professorship.

This research is based on observations made with the NASA/ESA Hubble Space Telescope obtained from the Space Telescope Science Institute, which is operated by the Association of Universities for Research in Astronomy, Inc., under NASA contract NAS 5–26555. These observations are associated with program(s) 15950.

This publication makes use of data products from the Wide-field Infrared Survey Explorer, which is a joint project of the University of California, Los Angeles, and the Jet Propulsion Laboratory/California Institute of Technology, funded by the National Aeronautics and Space Administration. This publication makes use of data products from the Two Micron All Sky Survey, which is a joint project of the University of Massachusetts and the Infrared Processing and Analysis Center/California Institute of Technology, funded by the National Aeronautics and Space Administration and the National Science Foundation. The Pan-STARRS1 Surveys (PS1) and the PS1 public science archive have been made possible through contributions by the Institute for Astronomy, the University of Hawaii, the Pan-STARRS Project Office, the Max-Planck Society and its participating institutes, the Max Planck Institute for Astronomy, Heidelberg and the Max Planck Institute for Extraterrestrial Physics, Garching, The Johns Hopkins University, Durham University, the University of Edinburgh, the Queen's University Belfast, the Harvard-Smithsonian Center for Astrophysics, the Las Cumbres Observatory Global Telescope Network Incorporated, the National Central University of Taiwan, the Space Telescope Science Institute, the National Aeronautics and Space Administration under Grant No. NNX08AR22G issued through the Planetary Science Division of the NASA Science Mission Directorate, the National Science Foundation Grant No. AST-1238877, the University of Maryland, Eotvos Lorand University (ELTE), the Los Alamos National Laboratory, and the Gordon and Betty Moore Foundation. 

This work has made use of data from the European Space Agency (ESA) mission {\it Gaia} (\url{https://www.cosmos.esa.int/Gaia}), processed by the {\it Gaia} Data Processing and Analysis Consortium (DPAC, \url{https://www.cosmos.esa.int/web/Gaia/dpac/consortium}). Funding for the DPAC has been provided by national institutions, in particular the institutions participating in the {\it Gaia} Multilateral Agreement.  

This research has made use of the SVO Filter Profile Service (\url{http://svo2.cab.inta-csic.es/theory/fps/}) supported from the Spanish MINECO through grant AyA2014-55216. This research has made use of Astropy, a community-developed core Python package for Astronomy \citep{astropy}.

\bibliographystyle{yahapj.bst}

\end{document}